\documentclass[prl,twocolumn,showpacs,preprintnumbers,letterpaper]{revtex4}
\usepackage{graphicx}
\usepackage{amsmath}
\usepackage{epsfig}
\usepackage{amssymb}

\begin{document}
% braket.sty          Macros for Dirac bra-ket <|> notation and sets {|}
% Donald Arseneau     asnd@triumf.ca     Last modified 05-Dec-1999.
% This is free, unencumbered, unsupported software.
%
% Commands defined are:
% \bra{ }   \ket{ }   \braket{ }   \set{ }    (small versions)
% \Bra{ }   \Ket{ }   \Braket{ }   \Set{ }    (expanding versions)
%
% The "small versions" use fixed-size brackets independent of their
% contents, whereas the "expanding versions" make the brackets and
% vertical lines expand to envelop their contents (internally using
% the \left and \right commands).  You should use the vertical bar
% character "|" to input any extra vertical lines.  In \Braket these
% vertical lines will expand to match the arguments, and in \Set the
% first vertical will expand this way.  E.g.,
%   \Braket{ \phi | \frac{\partial^2}{\partial t^2} | \psi }
%   \Set{ x\in\mathbf{R} | 0<{|x|}<5 }
%
% NOT defined is "\ketbra" (for projection operators) because I prefer
% \ket{ } \bra{ }.
%
% Because each definition is so small, it makes no sense to have a
% complicated generic version for many bracket styles.  Instead,
% you can just copy the definitions and change \langle or \rangle,
% < and > to what you like.
%
\def\bra#1{\mathinner{\langle{#1}|}}
\def\ket#1{\mathinner{|{#1}\rangle}}
\def\braket#1{\mathinner{\langle{#1}\rangle}}
\def\Bra#1{\left<#1\right|}
\def\Ket#1{\left|#1\right>}
{\catcode`\|=\active
  \gdef\Braket#1{\left<\mathcode`\|"8000\let|\BraVert {#1}\right>}}
\def\BraVert{\egroup\,\mid@vertical\,\bgroup}
% The \mid@vertical is \vrule with ordinary TeX but \middle| in eTeX.
% We always avoid a \mathchoice in making the inner vertical lines.
% Note that \right>, prints the same as \right\rangle but is faster.
%
% \def\ketbra#1#2{\ket{#1}\bra{#2}}
% \def\Ketbra#1#2{\left|{#1}\vphantom{#2}\right>\left<{#2}\vphantom{#1}\right|}

% \Set{...|...} Only the first | is treated specially.
{\catcode`\|=\active
  \gdef\set#1{\mathinner{\lbrace\,{\mathcode`\|"8000\let|\midvert #1}\,\rbrace}}
  \gdef\Set#1{\left\{\:{\mathcode`\|"8000\let|\SetVert #1}\:\right\}}}
\def\midvert{\egroup\mid\bgroup}
\def\SetVert{\egroup\;\mid@vertical\;\bgroup}

% If the user is using e-TeX with its \middle primitive, use that for
% verticals instead of \vrule.
%

\title{Spectrum and Dynamics of the BCS-BEC Crossover from a Few-Body Perspective.}
\author{Javier von Stecher}
\author{Chris H. Greene}
\affiliation{Department of Physics and JILA, University of Colorado, Boulder, Colorado
80309-0440}

\begin{abstract}
The spectrum of two spin-up and two spin-down fermions in a trap is
calculated using a correlated gaussian basis throughout the range of
the BCS-BEC crossover. These accurate calculations provide a
few-body solution to the crossover problem. This solution is used to
study the time-evolution of the system as the scattering length is
changed, mimicking experiments with Fermi gases near Fano-Feshbach
resonances. The structure of avoiding crossings in the spectrum
allow us to understand the dynamics of the system as a sequence of
Landau-Zener transitions. Finally, we propose a ramping scheme to
study atom-molecule coherence.
\end{abstract}

\maketitle

Optical lattices are a powerful tool to study few body systems. When
tunneling is negligible, optical lattices can be viewed as an
ensemble of individual harmonic traps where the properties of these
systems can be studied. The interaction between the particles can be
tuned using a Fano-Feshbach resonance \cite{FRes} and the number of
particles in each lattice site can be controlled
\cite{Esslinger,miroshnychenko}. In a recent experiment with optical
lattices, the spectrum of two fermions in a trap has been measured
\cite{Esslinger}, demonstrating that few body trapped systems can be
studied in their own right. Also, the BCS-BEC crossover has been
routinely explored in experiments with ultracold Fermi gases
\cite{regal2003cum,jochim2003bec,zwierlein2003obe,strecker2003caf,
bourdel2004esb,kinast2004esr}. In this Letter, we explore the
spectrum and dynamics of four trapped particles and we show how a
few-body formulation allows us to obtain accurate solutions of the
system without making the standard approximations of many-body
theory. This provides an explicit representation of avoided
crossings between the atomic degenerate Fermi gas (DFG or BCS)-type
states and molecular BEC-type states. Our results directly apply to
optical lattice experiments and they provide a few-body perspective
on BCS-BEC crossover dynamics.

Specifically, we calculate the spectrum of two pairs of trapped
fermionic atoms interacting through short-range potentials, all with
the same mass $m$. One pair is assumed to be distinguishable from
the other pair, but the two atoms within each pair are
indistinguishable. The s-wave scattering length $a$ of the
short-range interactions will be tuned in the standard
manner,\cite{FRes} which allows us to explore the BCS-BEC crossover
as a function of interaction strength near a broad Fano-Feshbach
resonance. Even though the BCS theory is not expected to apply to a
4-particle system, we still use this term to refer to the dynamical
regime where $a$ is small and negative. By solving the problem from
a few-body perspective, we are able to give accurate properties -
especially energy levels as well as time-dependent dynamics - of the
full quantum mechanical spectrum at zero temperature. As a result we
achieve a deeper understanding of the global topology of the
spectrum, in addition to making quantitative predictions of
transition probabilities and dynamical properties of this system
when interactions change with time as in experiments
\cite{regal2003cum,jochim2003bec,zwierlein2003obe,strecker2003caf,
bourdel2004esb,kinast2004esr}.

To obtain the energy spectrum, we use a correlated gaussian basis
set \cite{singer1960uge,varga1997sfb}. A diabatization procedure
reduces the system to a tractable number of relevant eigenfunctions,
after which we solve the time-dependent Schr\"{o}dinger equation
using the diabatic representation.

The Hamiltonian adopted is
\begin{equation}
\mathcal{H}=\sum_{i}^{4}\left( -\frac{\hbar ^{2}}{2m}\nabla _{i}^{2}+\frac{1%
}{2}m\omega _{0}^{2}\mathbf{r}_{i}^{2}\right)
+\sum_{i=1}^2\sum_{j=3}^4 V(\mathbf{r}_{ij}). \label{Ham}
\end{equation}%
In this convention, particles 1 and 2 are ``spin up" and particles 3
and 4 are ``spin down". The two-body potential function
$V(\mathbf{r}_{ij})$ is taken to be a purely attractive gaussian,
$V(\mathbf{r}_{ij})=V_{0}\exp(-\mathbf{r}_{ij}^{2}/2d_{0}^{2})$
where the width $d_{0}$ of the gaussian is fixed and the depth
$V_{0}$ is tuned to produce the desired two-body scattering length
$a$. To obtain results independent of the model potential
properties, we concentrate on the range $d_{0}\ll a_{ho}$, where
$a_{ho}=\left( \hbar /m\omega
_{0}\right) ^{1/2}$ is the trap length. By considering different widths $%
d_{0}$ ranging from $0.05a_{ho}$ to $0.01a_{ho}$ we have verified
that our results exhibit a weak dependence on $d_{0}$. All results
presented in this Letter correspond to $d_{0}=0.01a_{ho}$. The
eigenspectrum of Eq.(\ref{Ham}) is obtained by an expansion into
correlated Gaussian basis set, i.e.,
\begin{equation}
\Psi _{\{d_{ij}\}}(\mathbf{r}_{1},\mathbf{r}_{2},\mathbf{r}_{3},%
\mathbf{r}_{4})=\mathcal{S}\left\{ \psi _{0}(\mathbf{R}%
_{CM})e^{-\sum_{j>i}r_{ij}^{2}/2d_{ij}^{2}}\right\}  \label{Basis}
\end{equation}%
with
$\mathbf{R}_{CM}=(\mathbf{r}_{1}+\mathbf{r}_{2}+\mathbf{r}_{3}+\mathbf{r}_{4})/4$
the center-of-mass coordinate, $\psi _{0}$
the center-of-mass ground state $\psi _{0}(\mathbf{R}%
_{CM})=e^{-2R_{CM}^{2}/a_{ho}^{2}}$, and $\mathcal{S}$ the symmetrization
operator. Opposite-spin fermions are treated as distinguishable particles,
so $\mathcal{S}=(1-\mathcal{P}_{1,2})(1-\mathcal{P}_{3,4})$ where $%
\mathcal{P}$ is the permutation operator. The wavefunctions obtained
from this basis set are in the ground center-of-mass state (with
$J_{CM}=0$). The relative coordinate wavefunction has quantum
numbers $J^{\pi }=0^{+}$, since the basis functions only depend on
$\mathbf{R}_{CM}$ and the interparticle distances, $r_{ij}$.

The basis functions, defined in Eq.(\ref{Basis}), are characterized
by the set of values $\{d_{ij}\}$ which are selected semi-randomly.
While the $d_{ij}$ corresponding to different spin fermions are
selected to range from a fraction of $d_{0}$ up to a couple of times
$a_{ho}$ to describe dimer formation, those $d_{ij}$ corresponding
to
equal spin fermions are selected to be of the order of the trap length $%
a_{ho}$. The typical size of the basis set used in these
calculations is about 7000-15000. The advantage of the correlated
gaussian basis set is that all the matrix elements can be calculated
analytically in terms of the $\{d_{ij}\}$'s and the properties of
the two-body potential and the trap. Prior to diagonalizing the
Hamiltonian, a linear transformation eliminates linear dependence,
typically reducing the basis set size by less than $10\%$. The basis
set is fixed while $V_{0}$ is tuned to give different scattering
lengths, whereby matrix elements are calculated once and then used
to obtain the spectrum throughout the entire range of the BCS-BEC
crossover.

The accuracy and convergence of the calculations have been verified
in detail. The ground state energy agrees with fixed-node diffusion
Monte Carlo (FN-DMC) calculations throughout the BCS-BEC crossover
\cite{vonstechtbp}. For example, the ground state energy at
unitarity varies from $5.099\hbar\omega$ to $5.027\hbar\omega$ as we
change $d_0$ from $0.05a_{ho}$ to $0.01a_{ho}$; FN-DMC calculations
for a square well potential of range $0.01a_{ho}$ lead to a ground
state energy $5.069(9)\hbar\omega$ \cite{vonstechtbp}. The
difference between our results and the FN-DMC results is thus about
1\%, where the shape and range of the interactions start to play a
role. Higher excited states are in agreement with the BCS and BEC
limiting behaviors. Furthermore, the ground and excited states in
the BEC can be used to extract the dimer-dimer scattering length,
which agrees with the Petrov \textit{et al.} prediction
\cite{petrov2004wbd}, and the corresponding effective range
\cite{vonstechtbp}.

\begin{figure}[h]
\includegraphics[scale=0.5]{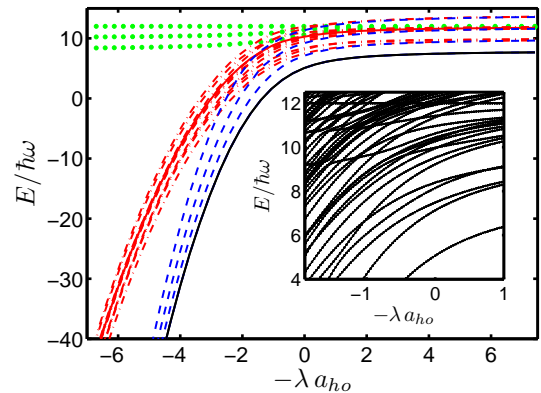}
\caption{(Color Online) The energy spectrum for four particles in a
trap is shown versus the dimensionless quantity $\protect\lambda
a_{ho}$ (20 states that correlate diabatically with the 20 lowest
energies in the noninteracting limit). The black solid curve is the
ground state. The blue dashed curves are states that diabatically
approach excited dimer-dimer configurations, the red dash-dotted
curves correspond to states that represent one dimer plus two free
atoms on the BEC side, while the states in green circles correlate
diabatically to four free atoms. The lowest curve drawn with green
circles is the Fermi gas ground state on the BEC side of the
resonance. Inset: zoom of the adiabatic spectrum in the crossover
regions. In this figure, all states are considered, showing the rich
structure of avoiding crossings.}\label{Spectrum}
\end{figure}

The spectrum as a function of $\lambda \equiv 1/a$ shows a series of
apparent crossings and avoided crossings in the unitarity region.
The avoided crossings can be roughly characterized by their width
$\Delta \lambda $, the range over which the two adiabatic
eigenstates interact appreciably, into two main categories: narrow
crossings, where $\Delta \lambda \ll 1/a_{ho}$, and wide crossings,
where $\Delta \lambda \gtrsim 1/a_{ho}$. We adopt a variant of the
diabatization procedure presented in Ref. \cite{HesseLin} to
diabatize narrow crossings while leaving wide crossings adiabatic,
which gives smooth energy curves. Figure~\ref{Spectrum} presents the
partially-diabatic spectrum in the BCS-BEC crossover. The inset
shows a zoom of the transition region, i.e., the strongly
interacting regime where the avoiding crossings occur. This
structure of avoiding crossings permits a global view of the manner
in which states evolve from weakly interacting fermions at $a<0$ to
all the different configurations of a Fermi gas at $a>0$, i.e.
molecular bosonic states, fermionic states and molecular boson-Fermi
mixture (see Fig.~\ref{Spectrum}). Furthermore, it allow us to
visualize concretely the possible pathways of the time-dependent
sweep experiments, as is shown below.

A partially-diabatic representation can be used to describe a ramp
of an initial configuration through the BCS-BEC crossover as in the
experiments carried out at different laboratories, like JILA and
Rice. The initial configuration is propagated using the time
dependent Schr\"{o}dinger equation. Starting from the ground state
in the BCS side, the parameter $\lambda $ is ramped through the
resonance to the BEC side at different speeds $\nu =\frac{d\lambda
}{dt}$. In a homogenous system, the ramps are only characterized by
the initial density $\rho$, the speed $\nu$ and $m$. This suggests
that $\chi \equiv \frac{m}{\hbar \rho }|\frac{d\lambda }{dt}|$ is
the relevant dimensionless quantity that characterize the ramp speed
in large systems. Thus, we will use $\chi$ to compare 4-body results
with large systems.  This idea of using the density, and not
properties of the trap, to connect few-body calculations with large
systems has been previously implemented \cite{borca2003tap}.

To interpret our numerical results we apply the Landau-Zener
approximation, which predicts that the probability for a transition
from the adiabatic $\Psi_j$ to $\Psi_i$ is $T_{ij}(\chi
)=e^{-\varkappa _{ij}/\chi }$, where $\varkappa _{ij}$ are
dimensionless parameters extracted from properties of the adiabatic
eigenstates, as is discussed below.

The nonadiabatic coupling or $P$-matrix controls the probability of
nonadiabatic transitions to a good approximation using the
Landau-Zener model. The coupling between two adiabatic states $\Psi
_{i}$ and $\Psi _{j}$ is $P_{ij}(\lambda )\equiv
\braket{\Psi_i|\frac{\partial\Psi_j}{\partial\lambda}}$ where
$\lambda $ is the adiabatic parameter. Clark has shown that, if the
transition has a form consistent with the Landau-Zener
approximation, then the $P$-matrix element for a transition from
$\Psi _{j}$ to $\Psi _{i}$ has a Lorentzian form whose width, along
with the corresponding eigenenergies, characterize the Landau-Zener
parameter $\varkappa _{ij}$ \cite{clarkpmatrix}. We numerically
evaluate all the potentially important $P_{ij}$ and verify that the
largest couplings generally have a smooth single-peak form that is
approximately Lorentzian. The largest $P_{ij}$ relevant for this
specific dynamical sweep correspond to transitions among $\Psi_1$,
$\Psi_2$, $\Psi_5$ and $\Psi_{14}$. Here, the partially-diabatic
states are labeled in increasing order of their energy on the BCS
side (see Fig. \ref{Spectrum}). $\Psi_1$ refers to the ground state,
$\Psi_2$ is the first excited dimer-dimer configuration, $\Psi_5$ is
the second configuration with a dimer and two atoms, and $\Psi_{14}$
is the lowest configuration with no dimers, the ``fermionic ground
state on the BEC side". The final probability distribution can be
explained as sequence of Landau-Zener transitions between these four
partially-diabatic states where the positions of the peaks of the
$P_{ij}(\lambda )$ determine a specific order in which the
transitions occur, namely $1\rightarrow 2$, followed by
$2\rightarrow 5$ and $ 1\rightarrow 5,$ and finally $5\rightarrow
14$.

\begin{figure}[h]
\includegraphics[scale=0.5]{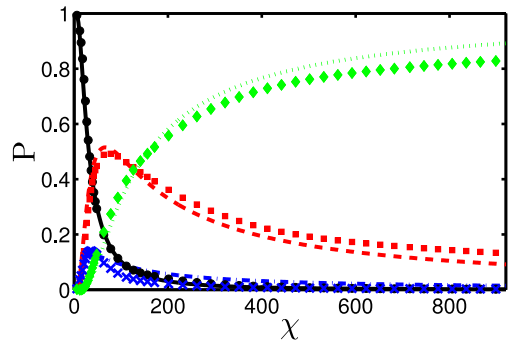}
\caption{(Color Online) The probability of evolving into a given
configuration is shown as a function of the dimensionless ramp speed
parameter $\protect\chi $. The symbols correspond to the full
numerical solution while the curves are Landau-Zener results. The
black solid curve and circles correspond to the ground state
configuration. The blue dash-dotted curve and crosses correspond to
higher dimer-dimer excitations. The red dashed curve and squares
correspond to ramps that produce a dimer plus two free atoms. The
green dotted curve and diamonds correspond to the lowest
configuration of four free atoms, i.e., the Fermi gas ``ground
state'' on the BEC side of the resonance.} \label{TransFig}
\end{figure}

We use $p_{i}$ to denote the probability of ending up in state $\Psi
_{i}$ following one ramp that started out in the DFG ground state
(\#1) on the BCS side. This Landau-Zener model then predicts that
\begin{gather}
p_{1}=(1-T_{5,1})(1-T_{2,1}),   \notag \\
p_{2}=(1-T_{5,2})T_{2,1},  \notag \\
p_{5}=(1-T_{14,5})\left[T_{5,1}(1-T_{2,1})+T_{5,2}T_{2,1}\right],  \label{probs}\\
p_{14}=T_{14,5}\left[T_{5,1}(1-T_{2,1})+T_{5,2}T_{2,1}\right].
\notag
\end{gather}%
The sum of all these probabilities is unity by construction. The
Landau-Zener parameters obtained from the P-matrix analysis are
$\varkappa _{2,1}\approx 35$, $\varkappa _{5,1}\approx 43$,
$\varkappa _{5,2}\approx 14$ and $\varkappa _{14,5}\approx 90$.
Interestingly, the Landau-Zener model shows decent agreement with
the numerical results despite the neglect of many possible
transitions.  Figure \ref{TransFig} presents the probability of
evolving in a certain configuration following a single
unidirectional ramp, as a function of the dimensionless ramp speed
parameter $\chi $. The numerical ramps are initiated at
$\lambda_i\sim -7/a_{ho}$ and finalized at $\lambda_f\sim 7/a_{ho}$.
As the speed is increased, the final state changes from a molecular
bosonic ground state, i.e., $\Psi_1$, to a \textquotedblleft
fermionic ground state\textquotedblright, i.e., $\Psi_{14}$, in
qualitative agreement with experiments.

To relate our results with experiments carried out at JILA and Rice,
we write the Landau-Zener parameter for atom-molecule transitions
$\delta=\varkappa_{mol}/\chi$ in terms of experimentally accessible
variables. If the dependence of $a$ on the magnetic field is
approximated in the usual manner as $a(B)=a_{bg}\left(
1+\frac{w}{B-B_{0}}\right) $ then $\nu=(dB/dt)/w a_{bg}$ (measured
near unitarity, i.e., where the transitions occur). Therefore,
$\delta=\varkappa_{mol} (dB/dt)^{-1} \rho \hbar \left\vert
wa_{bg}\right\vert /m$ which agrees with previous theoretical
predictions \cite{goral2004aau,williams2006tfm}. The dependence of $
\delta$ on $\rho$ has been experimentally verified
\cite{hodby2005peu}. To evaluate $\chi $ we use the average density
for the non-interacting 4-particles Fermi gas, namely $\rho \approx
0.153/a_{ho}^{3}$. The molecular fraction, the fraction of atoms
that become molecules after the ramp is over, is probably the most
relevant quantity to compare with experiments. For our 4-body
system, the molecular fraction is defined as the probability of
ending up in a dimer-dimer configuration plus half of the
probability to form a configuration of the ``dimer plus two free
atom" type. In the experiments we compare with, the molecule
fraction was fitted to a Landau-Zener function, i.e.
$p_{mol}^{LZ}=f_{m}(1-e^{-\varkappa _{mol}/\chi })$, where $f_{m}$
is is the maximum conversion efficiency which depends on
temperature.

 Whether a Landau-Zener function is the correct functional form to
describe the molecule formation fraction in large systems remains a
question that existing experiments have not resolved
\cite{pazy2005nap,altman2005dpf,williams2006tfm}. The Landau-Zener
model presented in this work for four particles does not predict a
single Landau-Zener function but a combination of different
Landau-Zener terms. However, the final molecule fraction predicted
by this model and the numerical results for the molecule formation
fraction can be approximately fitted by this Landau-Zener function
with $\varkappa_{mol}^{(4)}\approx 59\pm 6$; this value is higher
than the two-body prediction of $\varkappa_{mol}^{(2)}\approx 42$.
%In both calculations $f_{m}=1$ since we are at $T$=0.
Our results are consistent with the experimental Landau-Zener
parameter obtained in Ref.\cite{regal2003cum} for $^{40}$K. The fit
of the experimental data to a Landau-Zener formula predicted a
$\varkappa ^{exp}_{mol}\approx 62\pm 15$. Also, experiments carried
out at Rice measured the Landau-Zener parameter for $^{6}$Li
\cite{strecker2003caf}. Taking into account the conditions of the
experiment and the properties of the $^{6}$Li Fano-Feshbach
resonance at $B\approx 543.8G$, we estimate $\varkappa
^{exp}_{mol}\sim 90$. Both experiments were  carried out at finite
temperature and, consequently, maximum conversion efficiencies were
approximately $f_{m}\sim 0.5$, while our calculations are at $T=0$
where $f_{m}=1$. In summary, both experiments are in general
agreement with the four-body predictions. In the following, we
propose more sophisticated sweeps to study atom-molecule coherence.

\begin{figure}[h]
\includegraphics[scale=0.6]{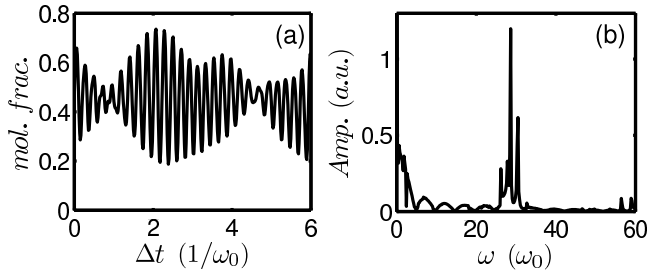}
\caption{(a) Molecular formation fraction is shown as a function of the delay $%
\Delta t$. (b) Fourier transform of the Fig.3(a). The peaks of the
spectrum correspond to the most important energy transitions. }
\label{molosc}
\end{figure}

Atom-molecule oscillations \cite{donley2002amc} or quantum beats
\cite{borca2003tap} have previously been explored for condensates,
and also for fermionic systems near a narrow Fano-Feshbach
resonance\cite{andreev2004nda}. To study atom-molecule coherence in
a Fermi gas near a broad Fano-Feshbach resonance, we have considered
different ramping schemes and noticed one that enhances the
atom-molecule oscillations. Starting in the ground state on the BCS
side, one ramps at medium speed ($\chi \sim 80$) to the BEC side and
pauses for a time $\Delta t$ at a value $\lambda _{still}$; then one
ramps back at the same speed to the BCS side where the scattering
length is close to zero. Finally, one slowly ramps $\lambda $ to the
BEC ($\chi \sim 7$) side and measures the resulting molecular
fraction. This is shown as a function of $\Delta t$ in Fig.
\ref{molosc}(a). Observe that this ramping scheme produces large
coherent oscillations in the molecular fraction. To interpret their
frequencies we Fourier transform the time-dependent molecular
fraction (see Fig. \ref{molosc}(b)). The frequency domain peaks
correlate with the most important configurations during the waiting
period at $\lambda _{still}\sim 5/a_{ho}$. The Bohr frequencies at
$\omega \approx 28\omega _{0}$ correspond to coherences between
states differing in one broken dimer bond. For example, the highest
peak is a coherence between $\Psi _{5}$ and $\Psi _{14}$ while the
second highest is a coherence between $\Psi _{1}$ and $\Psi _{5}$.
The frequencies around $\omega \approx 57\omega _{0}$ correspond to
coherences between states differing in two broken dimers bonds, e.g.
coherences between $\Psi _{1}$ and $\Psi _{14}$, and between $\Psi
_{2}$ and $\Psi _{14}$. In optical lattice experiments, this kind of
multipeak structure should be particularly pronounced in a tight
trap.

The four body problem remains fundamental and challenging. We have
presented an accurate numerical solution of the spectrum of two
\textquotedblleft spin up\textquotedblright\ and two
\textquotedblleft spin down\textquotedblright\ fermions in a trap
throughout the range of the BCS-BEC crossover. Even though the
spectrum presents a rich structure of avoided crossings, we have
shown that a simple Landau-Zener model approximately describes the
dynamics of unidirectional ramps. The spectrum and dynamics of this
system is interesting for optical lattice experiments. These would
allow access to physics that cannot be probed in the two-body
system, like corrections to the energy spectrum due to the
dimer-dimer interaction, and also to the atom-dimer interaction.
Also, the system of two \textquotedblleft spin up\textquotedblright\
and two \textquotedblleft spin down\textquotedblright\ fermions in a
trap exhibits many of the ingredients of the BCS-BEC crossover
problem, and in that sense the present results provide a few-body
perspective on Fermi gas experiments.

We are grateful to Doerte Blume for providing unpublished data to
us, and Josh Dunn, Seth Rittenhouse, Nirav Mehta and Jose D'Incao
for useful discussions. This work was supported in part by NSF.

\end{document}